# A new material for probing spin-orbit coupling in Iridates

**Brendan F. Phelan, Jason Krizan, Weiwei Xie, Quinn Gibson, and R. J. Cava**

Department of Chemistry, Princeton University, Princeton NJ 08540

ABSTRACT

We report the structure and magnetic properties of a new iridate compound, $Sr_xLa_{11-x}Ir_4O_{24}$, where the $d$-electron count of Ir and therefore its number of unpaired electrons can be tuned continuously from $5d^5$ $Ir^{4+}$ to $5d^4$ $Ir^{5+}$, i.e. from $SrLa_{10}Ir_4O_{24}$ to $Sr_5La_6Ir_4O_{24}$. The $IrO_6$ octahedra in $Sr_xLa_{11-x}Ir_4O_{24}$ are isolated from each other and from other transition elements, minimizing band effects, and the doping is on the framework sites, not the Ir sites, minimizing the effects of disorder. Measurements of the temperature dependent magnetic susceptibility are employed to determine the evolution of the Ir magnetic moment on progressing from $5d^5$ $Ir^{4+}$ to $5d^4$ $Ir^{5+}$, and are clearly best described by a transition from a $J=1/2$ to a $J=0$ Ir magnetic state; that is, the evolution of the magnetic susceptibility shows the dominance of spin-orbit coupling in determining the magnetic properties of a material with highly isolated $IrO_6$ octahedra.

I. INTRODUCTION

Spin Orbit Coupling (SOC) has recently been of significant interest in advancing the understanding of correlated electron materials. Its consequences appear to be manifested in materials as diverse as oxides, superconductors, and topological insulators (1-12). Iridium oxides are an important class of such materials, and have become the topic of many recent studies both experimentally and theoretically. Of particular interest have been systems based on $Ir^{4+}$ in $IrO_6$ octahedra, where the electron configuration is $[Xe]5d^5$. The octahedral crystal field places the five $d$ electrons in the $t_{2g}$ $d$-state manifold, which, for ideal octahedra, is triply degenerate. It has been widely discussed that strong SOC in $Ir^{4+}$ systems will split the $t_{2g}$ manifold into two effective angular momentum energy levels: a doublet $J_{eff} = 1/2$ and a quartet $J_{eff} = 3/2$. For $Ir^{4+}$, the $J_{eff} = 3/2$ bands are completely filled while the $J_{eff} = 1/2$ bands are half filled, giving rise to a J = 1/2 system with one unpaired electron in a two electron band (see e.g. refs. 1 and 2). Studies by techniques such as resonant inelastic X-ray scattering have shown that the real case for individual materials can often be more complex due to the presence of crystal field splitting, which can also lift the $t_{2g}$ degeneracy (see e.g. refs. 13-20).

An alternative to studying the detailed properties of an iridate with a one Ir $d$-electron configuration would be to probe the importance of SOC in a material where the systematic tuning of $d$-state electron population is possible and the effect of that change on the magnetism can be characterized. Unfortunately, for the materials studied thus far, systematic tuning over a wide range of Ir $d$-electron counts is difficult (see. e.g. ref. 21). Here we present a new iridium oxide that due to the characteristics of its structure and its doping flexibility provides straightforward support for the importance of SOC in iridium oxides through measurements of its magnetic susceptibility as its iridium is continuously tuned between $5d^5$ $Ir^{4+}$ and $5d^4$ $Ir^{5+}$. This material has the general formula $Sr_xLa_{11-x}Ir_4O_{24}$ and features an array of isolated $IrO_6$ octahedra. The unique combination of the grossly insulating character of the material, the absence of other transition metals, and the weak magnetic interactions between neighboring octahedra result in a good model material for examining SOC in iridates. These characteristics simplify the interpretation of the magnetic data by effectively eliminating the complexities that may arise due to band effects and strong magnetic interactions. Our ability to change the value of $x$ between 1 and 5 in $Sr_xLa_{11-x}Ir_4O_{24}$ allows for the continuous tuning of the iridium oxidation state from $Ir^{4+}$ to $Ir^{5+}$ that is critical for proving the point: $5d^5$ $Ir^{4+}$ in an ideal octahedron will display the behavior of a single unpaired electron in its magnetic susceptibility, regardless of whether SOC is present or not. Conversely, $5d^4$ $Ir^{5+}$ will either have no unpaired electrons when strong SOC is present, or two unpaired electrons when SOC is absent. This substantial difference allows us to make an assessment of the importance of SOC in iridium oxide octahedra simply through measurements of the magnetic susceptibility, supported by an estimate of the influence of crystal field effects through electronic structure calculations. If the effective moment ($\mu_{eff}$) per iridium had increased on going from $Ir^{4+}$ to $Ir^{5+}$, we would have concluded that SOC is relatively weak. However, as is the actual case in $Sr_xLa_{11-x}Ir_4O_{24}$, a strong decrease in $\mu_{eff}$ on going from $Ir^{4+}$ to $Ir^{5+}$ indicates that SOC plays a dominant role in the magnetic behavior.

## II. EXPERIMENTAL

Small black single crystals of one of the material's compositions, $Sr_{4.25}La_{6.75}Ir_4O_{24}$, used to determine the crystal structure, were grown by heating a 1:4 mixture of $La_2O_3$ and $SrIrO_3$ in a $SrCl_3$ flux with a ratio of 7 to 1 flux to sample. After drying at 250 °C, the mixture was sealed in a platinum tube by crimping and welding the ends of the tube. The loaded tube was first heated to 1000 °C at a rate of 3 °C/min then heated to 1260 °C at a rate of 1.5 °C/min where it was held for 94 h. The tube was cooled slowly at 3.5 °C/h to 850 °C to allow crystal growth. The sample was then cooled to room temperature at 3 °C/min. After washing away excess $SrCl_2$ with deionized water, single crystals of $Sr_{4.25}La_{6.75}Ir_4O_{24}$ of suitable size and quality for single crystal X-ray diffraction were obtained. Single crystal diffraction data was collected on Bruker APEX II with graphite monochromated Mo K$\alpha$ $\lambda$=0.713 Å at 20 °C. Data collection utilized the Bruker APEXII software package and subsequent reduction and cell refinement were performed using Bruker SAINT (22). The crystal structure was determined through the use of SHELXL-2013 as implemented through the WinGX software suite (23, 24). The composition was determined through refinement of the Sr and La partial occupancies. This composition is in good agreement with the composition of crystals as determined by energy dispersive X-ray spectroscopy (EDS) implemented in a Quanta scanning electron microscope.

    Polycrystalline samples for the study of the magnetism of $Sr_xLa_{11-x}Ir_4O_{24}$ ($1 \leq x \leq 5$, $\Delta x = 0.5$) were prepared by solid state synthetic methods. Powdered samples were prepared from dried $La_2O_3$ (powder, 99.99%, Alfa Aesar), $SrCO_3$ (powder, 99.99%, Alfa Aesar), and Ir (powder, 99.95%, Alfa Aesar). Mixtures of starting materials were homogenized using a mortar and pestle before being pressed into ¼" pellets using a pellet die compacted in a hydraulic press. Pellets were heated at 1100 °C for 12 hours. Samples were reground and pelletized and heated again to 1100 °C for 12 hours. In order to

stabilize the desired oxidation states, samples where $1 \leq x \leq 4.5$ were then ground and heated under flowing argon to 1000 °C for 1 hour. For x = 5 the sample was reground, placed in an alumina crucible, sealed in a quartz ampule under positive oxygen pressure, and heated to 600 °C for 48 hours. Samples were characterized by powder X-ray diffraction (PXRD) using a Bruker D8 Focus diffractometer with Cu Kα radiation and a diffracted beam monochromator. Structural refinements of this diffraction data were processed using the FullProf software Suite, with peak shapes refined using the Thomson-Cox-Hasting pseudo-Voigt function with the background modeled as a Chebychev polynomial. (25, 26)

The oxygen stoichiometry of the materials was determined by thermogravimetric analysis (TGA) on a TA instruments SDT Q600 instrument. Samples were reduced in flowing $H_2$/Ar on heating to 700 °C at a rate of 0.5 °C/min. Reducing the samples by heating to 1000 °C at a rate of 1 °C/min under flowing Ar resulted in the desired oxygen content in samples where $1 \leq x \leq 4.5$. Temperature dependent magnetizations (M) were measured with a Quantum Design Magnetic Property Measurement System (MPMS). Measurements of the M vs. applied magnetic field H were linear up to applied fields of $\mu_0 H$ = 2 Tesla, and thus χ was defined as χ = $M/\mu_0 H$, at an applied field of 1 Tesla. Zero-field cooled (ZFC) measurements were performed on heating from 1.8 K to 250 K in a magnetic field of 1T. Resistivity measurements on polycrystalline pellets of all samples in this family at ambient temperature yielded very high values, greater than $2.5 \times 10^5$ ohm-cm. These resistivities indicate that the materials are highly insulating and therefore considering the $IrO_6$ octahedra to be electronically isolated is fully justified.

Electronic structure calculations on the hypothetical model compound "La$_{11}$Ir$_4$O$_{24}$" were carried out using the Vienna Ab-initio Simulation Package (VASP) with a plane wave cutoff energy of 500 eV, and a set of 6×6×4 *k* points for the irreducible Brillouin zone (27). Exchange and correlation were treated by the generalized gradient approximation (GGA). To describe the electron correlation associated with the La 4*f* states, on-site repulsion was applied using the LSDA+U method; the well-known Hubbard and exchange parameters of *U* = 6.7eV and *J* = 0.7eV for La 4*f* states were employed (28). To compare the effect of on-site repulsion for the Ir, LSDA with and without U were calculated, with *U* = 4.7eV and *J* = 0.7eV (28,29). Density of states calculations for the model Ir$^{4+}$ and Ir$^{5+}$ compounds SrLa$_{10}$Ir$_4$O$_{24}$ and Sr$_5$La$_6$Ir$_4$O$_{24}$ were performed using the experimentally determined structures by Tight-binding Linear-Muffin-Tin-Orbital Atomic Sphere Approximation (TB-LMTO-ASA) using the Stuttgart code (30). Exchange and correlation were treated by the local density approximation (LDA) and the local spin-density approximation (LSDA) (31). In the ASA method, space was filled with overlapping Wigner−Seitz (WS) spheres (32), and a combined correction is used to take into account the overlapping part. Empty spheres are necessary for this structure type, and the overlap of WS spheres is limited to no larger than 16%. To test the accuracy of the LMTO method for calculating magnetic moments for these materials, we performed calculations for hypothetical "La$_{11}$Ir$_4$O$_{24}$" using both the LMTO and VASP codes. The magnetic moments generated from the two methods are exactly the same. Moreover, the inclusion of on-site repulsion on Ir (with and without *U* = 4.7eV, and *J* = 0.7eV) does not affect the magnetic moments calculated. Based on this, considering the fact that the use of VASP for structures with large numbers of atoms per cell (184 in this material) leads to difficulties, the predicted magnetic moments for the representative intermediate members of the $Sr_xLa_{11-x}Ir_4O_{24}$ solid solution in the absence of spin orbit coupling were calculated by LMTO only.

## III. RESULTS AND DISCUSSION

$Sr_xLa_{10-x}Ir_4O_{24}$ has a previously unreported crystal structure for an iridate, the details of which are presented in Tables S1-S3. This crystal structure has been previously realized in the Re-based compound $Sr_{11}Re_4O_{24}$ (33). A key feature of this structure type from the magnetic perspective is the three-dimensional array of isolated metal-oxygen octahedra (Figure 1a and b). In the case of $Sr_xLa_{10-x}Ir_4O_{24}$ these $IrO_6$ octahedra are approximately 5.8 Å apart, with no intervening transition metals, and only slightly distorted (Figure 1c). The Ir-O bonds in the octahedra vary from the average bond length by ± 2.5% and the bond angles vary from 90 degrees by ± 7 degrees. These variations are comparable to those observed in other well-studied iridate systems, but the octahedra are more distorted than those found in $Ca_4IrO_6$ (34). These crystal structure characteristics make the new material a good candidate for studying SOC in iridates, as described further below.

The suitability of this compound for exploring the effects of SOC was initially realized with the synthesis of single crystals of $Sr_{4.25}La_{6.75}Ir_4O_{24}$. By changing the Sr to La ratio of $Sr_xLa_{11-x}Ir_4O_{24}$ in bulk samples, we tuned the oxidation state of iridium between 4+ and 5+. The oxygen content of the synthesized materials was confirmed by reducing samples with 2.5% $H_2$: 97.5% Ar in TGA (Figure 2). This fully reduced the samples to $La_2O_3$, SrO, and Ir metal, allowing us to calculate the relative amount of oxygen in each sample. These calculations revealed that the samples of compositions 1 ≤ $x \leq 4.5$ were prone to incorporating additional oxygen into the structure. By heating the as-synthesized materials in a weakly reducing Ar atmosphere we were able to stabilize the oxygen content of these samples (Figure 2 inset) at the desired 24 per formula unit. As is the case for all transition metal oxides where the stoichiometries of highly electropositive elements and oxygen are well determined; the oxidation state of the transition metal, in this case iridium, is well defined by charge neutrality requirements (35). After the bulk samples were synthesized for $1 \leq x \leq 5$, Rietveld refinements of the powder patterns were conducted using the basic crystal parameters from the single crystal X-ray data (Figure 3). After refining the data for each sample, the tetragonal lattice parameters *a* and *c* were found to change linearly with *x* (Figure 2 inset), as expected by Vegard's Law.

In order to determine the effective moment ($\mu_{eff}$) per iridium, we measured the zero field cooled (ZFC) magnetization versus temperature at an applied field of 1 T from 2 to 250 K. The plot of magnetic susceptibility ($\chi$) versus temperature in Figure 4 shows a systematically increasing $\chi$ as *x* decreases. The presence of an antiferromagnetic (AFM) transition is seen at approximately 10 K in compositions where $x \leq 3.5$. Specific characterization of this transition is not examined in the current work. Values for the Curie constant (C) and Weiss temperature (θ) were extracted by fitting plots of inverse susceptibility ($1/(\chi - \chi_0)$) versus temperature (Figure 5). In Figure 6 we show a dimensionless plot of the magnetic susceptibility that scales temperature by the Curie Weiss theta and the observed inverse susceptibility by the Curie constant for each compound studied. This plot allows all the compounds in the $Sr_xLa_{11-x}Ir_4O_{24}$ solid solution family to be compared directly in spite of the variations in effective moment and Curie Weiss theta. The compounds are all observed to obey Curie Weiss behavior up to high values of $T/\theta$ with no significant deviations from that behavior until they display magnetic ordering at low temperatures, evidenced by the kinks in the curves. These kinks occur at temperatures larger than $T/\theta = 1$ in materials where $x \leq 3$, suggesting that a competition between antiferromagnetic and ferromagnetic coupling is present to yield the very small $\theta$s observed, since ordering at temperatures above $\theta$ is not expected in magnetic systems with a single dominant magnetic coupling (36). The inset, which shows the systematic variation of $\theta$ with composition (it first becomes more negative and then becomes more positive) further supports the conclusion that a balance of antiferromagnetic and ferromagnetic coupling must be present in this system.

In Figure 7, the observed $\mu_{eff}$ values determined from C, based on $\mu_{eff} = \sqrt{8C}$, are plotted (black squares) against $x$ in $Sr_xLa_{11-x}Ir_4O_{24}$. The observed $\mu_{eff}$ decreases as $x$ increases, correlating to the transition from $Ir^{4+}$ to $Ir^{5+}$ in the compound. The result is consistent with the expectation that strong SOC interactions will cause $Ir^{5+}$ to have all paired spins and a net $\mu_{eff} = 0$. The simple point-charge local electron scheme used to calculate the $x$ dependence of the expected moments for this scenario is shown in the inset to the figure, and the resulting behavior is shown as a dashed line. This is vastly different from the expected moment of a system with no SOC interactions, where $Ir^{5+}$ would have two unpaired spins, and a $\mu_{eff} = 2.82$; the expected behavior for this scenario is shown as a dot-dashed line in the figure. Our experimental data, shown as black squares, indicates an excellent correlation with the expected $\mu_{eff}$ in a system that is subject to strong SOC.

Of interest for a complete analysis of the data is a consideration of the potential influence of crystal field (CF) splitting, which can also impact the $t_{2g}$ energy level degeneracies; it has, for instance, been found to be a significant factor in $Sr_3CuIrO_6$ and $CaIrO_3$ for instance (14, 15). We therefore studied the expected magnetic state in the absence of spin orbit coupling, but with crystal field splitting allowed, through first-principle calculations of the electronic Density of States (DOS) for selected compositions of $Sr_xLa_{11-x}Ir_4O_{24}$. The structural models employed (Table S4) distribute the Sr and La s1distributions; the conclusion drawn about the magnetic behavior was independent of the selection of different sets of relative Sr/La site occupancies. The results of the DOS calculations in the local density approximation with no spin polarization and no spin orbit coupling are shown in Figure 8 for the $Ir^{4+}$ case, $SrLa_{10}Ir_4O_{24}$, and the $Ir^{5+}$ case, $Sr_5La_6Ir_4O_{24}$. A distinct narrow band (total width about 1.5 eV) is seen for both materials straddling $E_F$. (Between approximately -1.5 eV and +0.2 eV for $Ir^{4+}$ $SrLa_{10}Ir_4O_{24}$, and -1.0 and + 0.3 eV for $Ir^{5+}$ $Sr_5La_6Ir_4O_{24}$.) For both materials, integrating the DOS in this band yields 24 electrons per formula unit, as is expected for the 3 $t_{2g}$ states/Ir x 2 electrons/state x 4 Ir/formula unit. The band is filled with 20 electrons in the former case and 16 electrons in the latter case. The Fermi levels are located on relatively sharp peaks in the calculated DOS in both cases, indicating that the electronic structures are likely to be unstable with respect to magnetism. (The same is seen for calculations that include spin polarization (Figure S1).) The height of the peak at $E_F$ is easily seen to be larger for the $5d^4$ $Ir^{5+}$ compound than it is for the $5d^5$ $Ir^{4+}$ compound, indicating that in the absence of spin orbit coupling, but including any crystal field splitting that may be present, this compound would be *more* magnetic than the $Ir^{4+}$ case, exactly the opposite of what is observed experimentally.

The magnetic moments calculated in LSDA, based on the stoner model, where the magnetic moment is derived from the size of the DOS peak, show this behavior more quantitatively. The resulting calculated magnetic moments per Ir for all the intermediate compositions tested are presented as triangular points in Figure 7. The moments do not exactly follow the line expected for an Ir ion in a perfect isolated octahedron with no SOC, which we interpret to reflect that fact that some CF splitting is present, but they clearly increase rather than decrease with increasing $Ir^{5+}$ content. This result shows that it is not possible for CF splitting to explain the experimentally observed large decrease in magnetic moment per Ir on going from $5d^5$ to $5d^4$ Ir in this material. Thus the electronic-structure-based comparison to the behavior observed in Figure 7 confirms the simple conclusion: the reduced moment observed can only be explained through the presence of SOC.

## IV. CONCLUSION

We have presented the crystal structure and elementary magnetic properties of the new iridate compound $Sr_xLa_{11-x}Ir_4O_{24}$. This compound, which is based on isolated $IrO_6$ octahedra, can be chemically manipulated through a solid solution between Sr and La to tune the electronic state of iridium between $5d^5$ $Ir^{4+}$ and $5d^4$ $Ir^{5+}$ for $1 \leq x \leq 5$. Analysis of the temperature dependent magnetic susceptibility as a function of $x$, supported by electronic structure calculations, shows that the observed magnetism cannot be explained without invoking the presence of strong spin orbit coupling for octahedrally coordinated

Ir in oxidation states between 4+ and 5+. Thus we argue $Sr_xLa_{11-x}Ir_4O_{24}$ is a valuable system for probing the importance of SOC in iridates.

**Acknowledgements**

The synthesis, magnetic characterization, and theoretical modeling of this compound were supported by the ARO MURIs on thermoelectrics and topological insulators, grants FA9550-10-1-0553 and W911NF-12-1-0461. The DOE, through grant DEFG02-08ER56544, supported the single crystal X-ray diffraction.

**Figure Captions**

**Figure 1:** Crystal structure of $Sr_{4.25}La_{6.75}Ir_4O_{24}$ presented with views in the *ac* plane (a) and the *bc* plane (b). These projections highlight the isolated $IrO_6$ octahedra. The two iridium sites are presented in dark blue (Ir1) and light blue (Ir2). The ball and stick models (c) show the slight distortions of the two different types of octahedra.

**Figure 2:** Thermogravimetic analysis data for $Sr_xLa_{11-x}Ir_4O_{24}$ with $x = 1$. Reduction in 2.5 % $H_2$: 97.5% Ar resulted in a calculated oxygen content of 25.3 per formula unit for samples synthesized in air. Reducing the sample in 100% Ar (inset) resulted in a stabilized oxygen content of 24 per formula unit.

**Figure 3:** Rietveld refinement of $Sr_2La_9Ir_4O_{24}$. Observed X-ray pattern is shown in red, calculated in black, and the difference ($I_{obs}$-$I_{calc}$) in blue. Green tick marks are Bragg reflections for $Sr_2La_9Ir_4O_{24}$. The inset shows the change in lattice parameters *a* and *c* as a function of *x* for $Sr_xLa_{11-x}Ir_4O_{24}$.

**Figure 4:** Susceptibility ($\chi$) versus temperature measured in a 1 T field for $Sr_xLa_{11-x}Ir_4O_{24}$ where $x$=1, 1.5, 2, 2.5, 3, 3.5, 4, 4.5, and 5. An AFM ordering transition is present in samples where x ≤ 3.5. Inset shows expansion of data from 0-25 K.

**Figure 5:** Inverse susceptibility (1 /( $\chi$ - $\chi_0$)) versus temperature in a 1 T field for $Sr_xLa_{11-x}Ir_4O_{24}$ where $x$=1, 1.5, 2, 2.5, 3, 3.5, 4, 4.5, and 5. Inset shows an example M vs. H plot for $Sr_3La_8Ir_4O_{24}$.

**Figure 6:** Plot of $C/|\theta|(\chi-\chi_0)$ versus $T/|\theta|$ for $Sr_xLa_{11-x}Ir_4O_{24}$ where $x$=1, 1.5, 2, 2.5, 3, 3.5, 4, 4.5, and 5. This plot shows that these compounds exhibit good Curie-Weiss behavior at high temperatures that persists down to their magnetic ordering temperatures. The inset

shows the Weiss temperature (θ) versus $x$ for $Sr_xLa_{11-x}Ir_4O_{24}$. The dotted line is a guide to the eye.

**Figure 7:** Observed $\mu_{eff}$ per Ir versus $x$ for $Sr_xLa_{11-x}Ir_4O_{24}$ shown as black squares. The red dashed line represents expected values of $\mu_{eff}$ with strong SOC where $Ir^{4+}$ is J= ½ and $Ir^{5+}$ is J = 0. The blue dashed line represents expected values of $\mu_{eff}$ without SOC, where $Ir^{4+}$ is still spin ½ but $Ir^{5+}$ is spin 1, based on $\mu_{eff} = g\, S(S+1)^{1/2}$ assuming a $g$ value of 2. The orange triangles represent the total Ir moment calculated from LMTO using LSDA with no SOC but including crystal field effects. The dotted line is a guide to the eye. The insert shows the splitting of the Ir $t_{2g}$ manifold into $J_{eff}$ = 1/2 and $J_{eff}$ = 3/2 states under strong SOC, resulting in J = ½ $Ir^{4+}$ and J = 0 $Ir^{5+}$.

**Figure 8:** The density of electronic states of $SrLa_{10}Ir_4O_{24}$ and $Sr_5La_6Ir_4O_{24}$ calculated by the LMTO method. The structural models presented in Table S3 were employed.

**Figure 1**

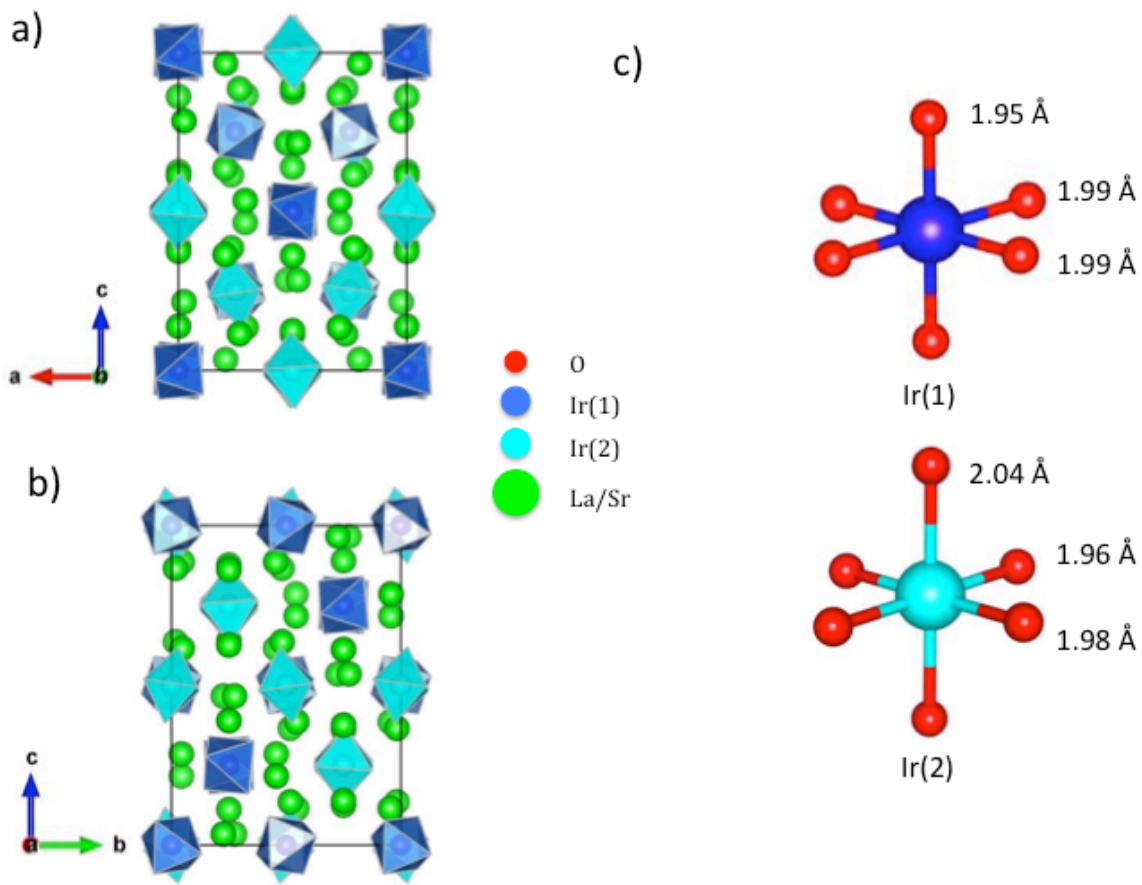

**Figure 2**

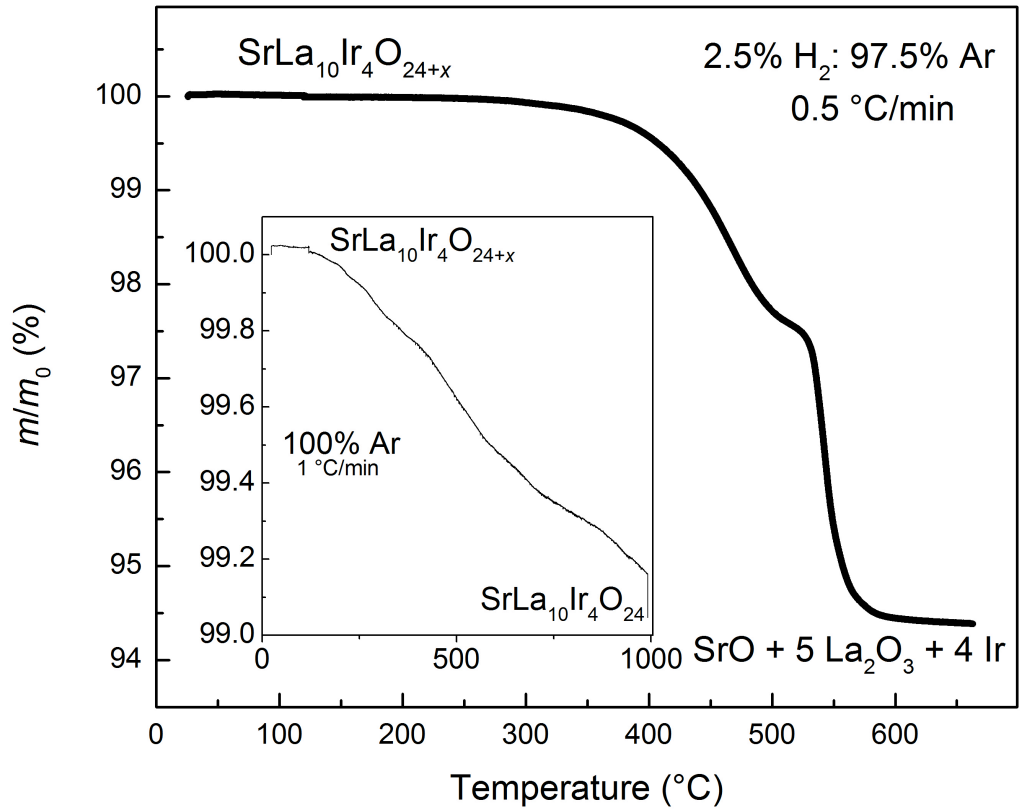

**Figure 3**

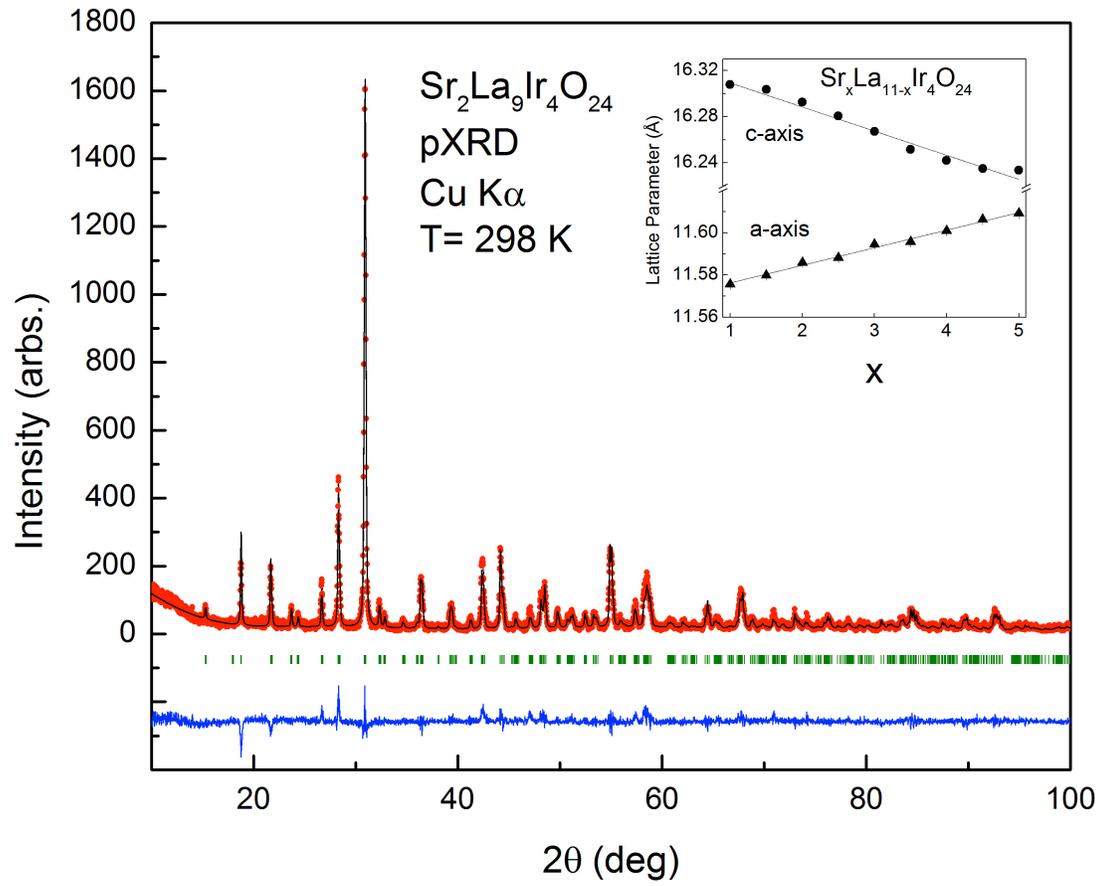

**Figure 4**

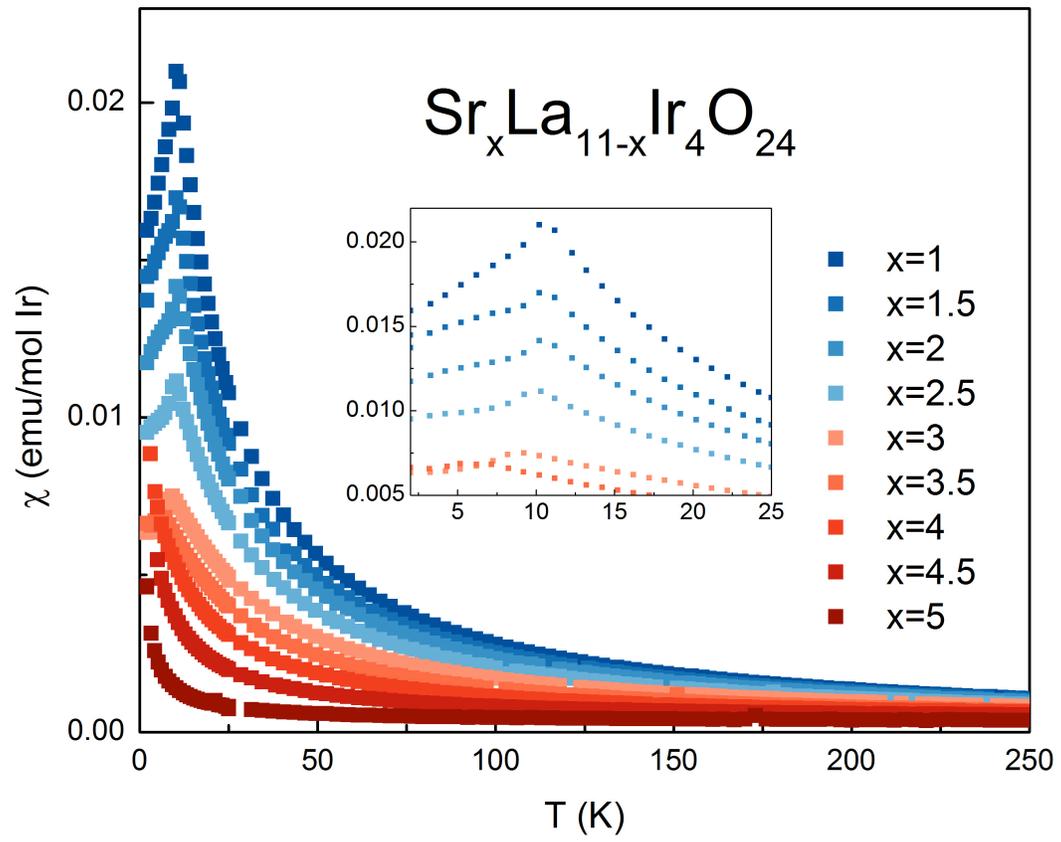

Figure 5

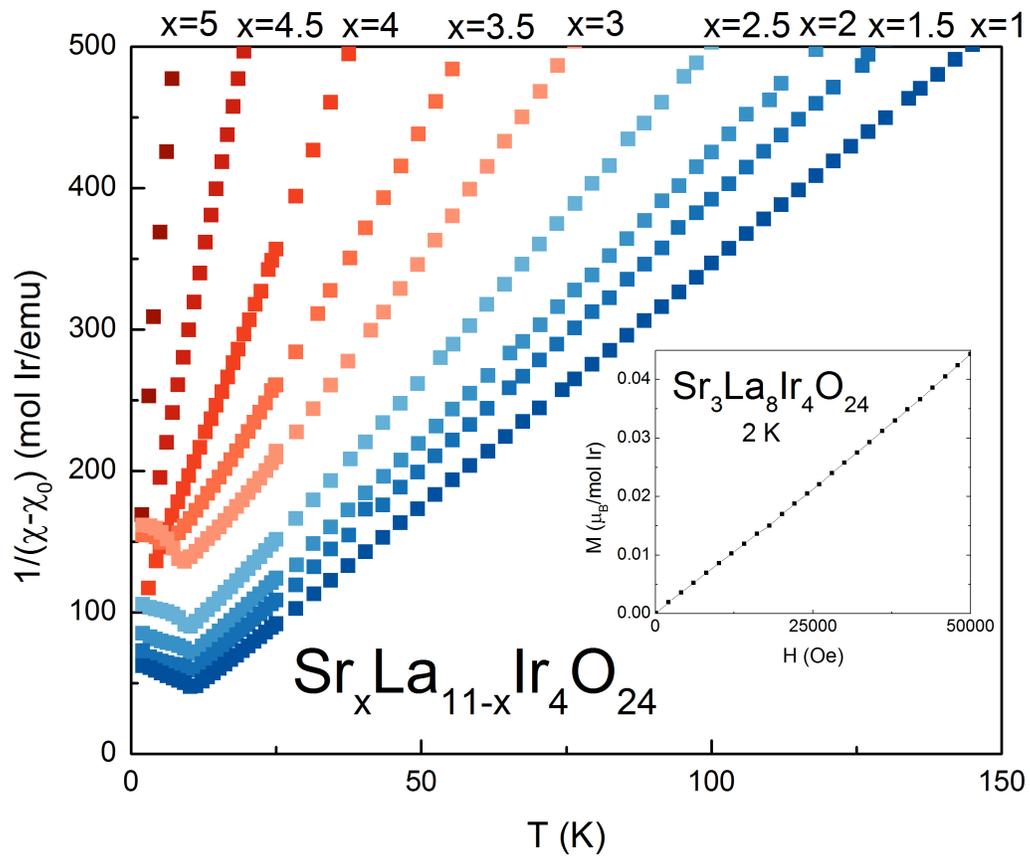

**Figure 6**

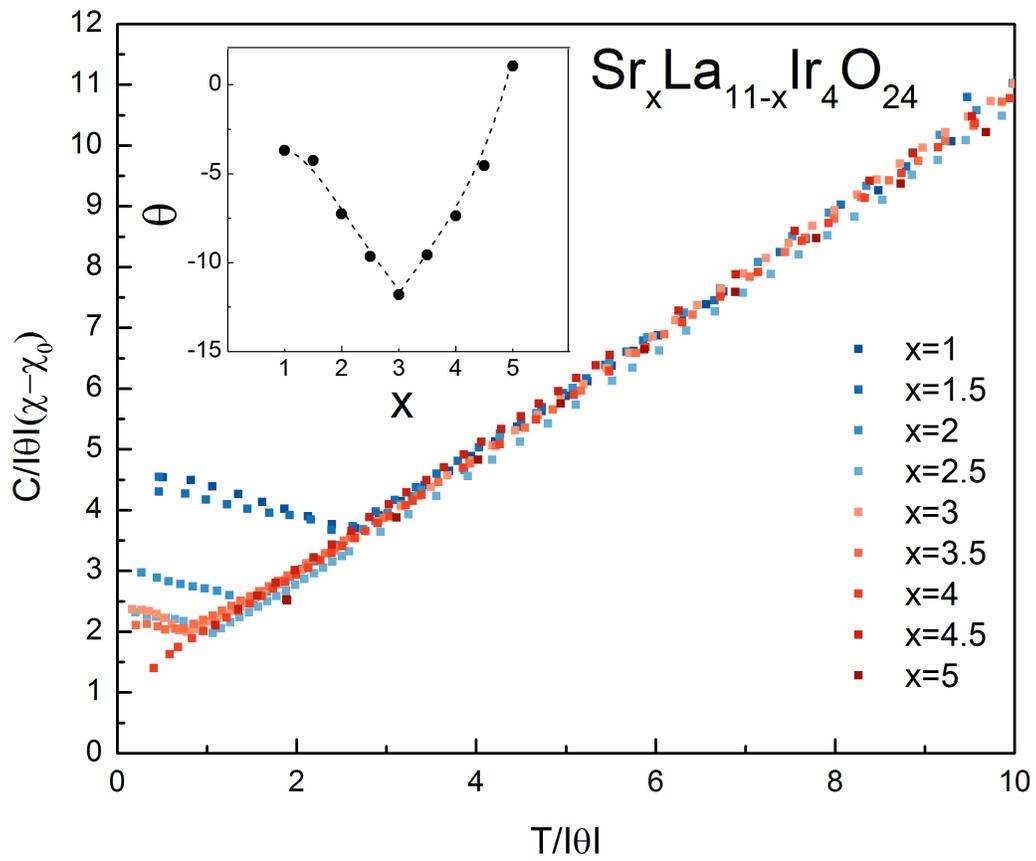

**Figure 7**

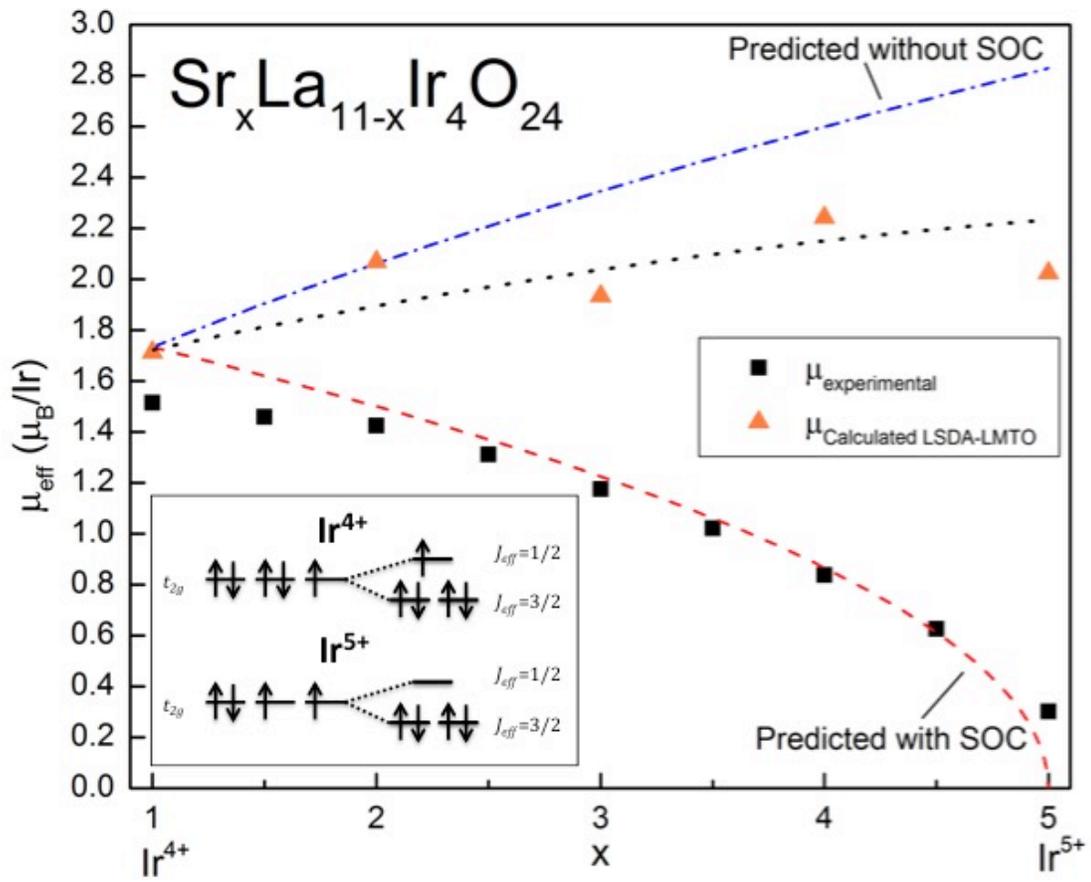

**Figure 8**

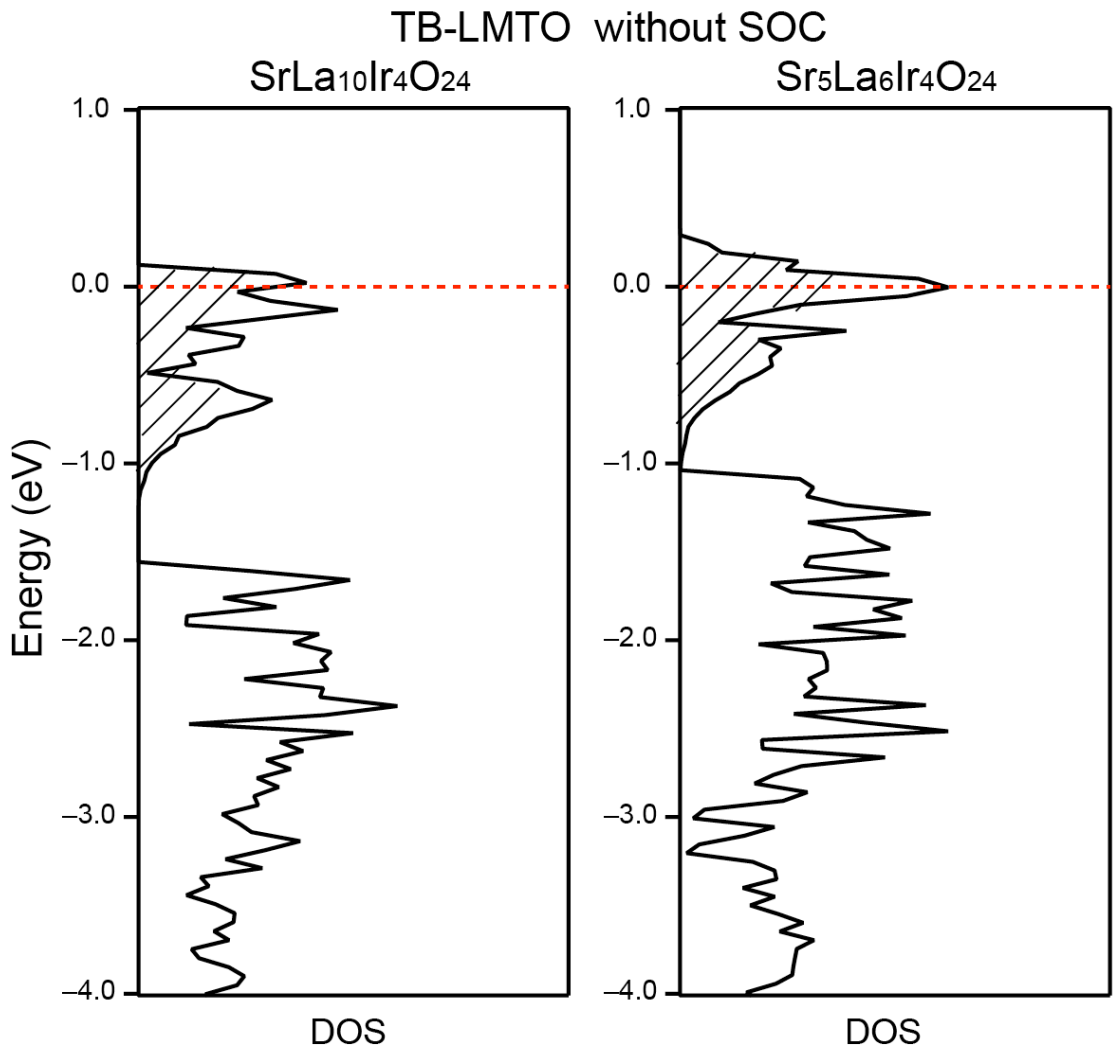

# Supplementary information for: A new compound for testing Spin-Orbit Coupling in Iridates

**Brendan F. Phelan, Jason Krizan, Weiwei Xie, Quinn Gibson, and R. J. Cava**

**Department of Chemistry, Princeton University, Princeton NJ 08540**

**Table S1: Single Crystal Data and Structural Refinement for $Sr_{4.24}La_{6.76}Ir_4O_{24}$**

| | |
|---|---|
| Formula weight | 2462.84 g/mol |
| Crystal System | Tetragonal |
| Space Group | I 41/a (88) |
| Unit Cell | a= 11.6213(12) Å<br>c= 16.2001(17) Å |
| Volume | 2187.9(5) Å$^3$ |
| Z | 4 |
| Radiation | Mo Kα |
| T | 293 K |
| Absorption Coefficient | 47.439 |
| F(000) | 4185 |
| Reflections collected/unique | 12674/ 1338 $R_{int}$ = 0.0380 |
| Data/Parameters | 1338/67 |
| Goodness-of-fit | 1.078 |
| Final R indices [I>2σ(I)] | $R_1$=0.0236 , $wR_2$=0.0362 |
| Largest diff. peak and hole | 1.391 and -2.092 e A$^{-3}$ |

**Table S2. Structural parameters and equivalent isotropic thermal displacement parameters for $Sr_{4.24}La_{6.76}Ir_4O_{24}$.**

| Atom | Wyck. | S.O.F. | x/a | y/b | z/c | U [Å2] |
|---|---|---|---|---|---|---|
| O1 | 16f | 1 | -0.1013(3) | -0.1319(3) | 0.0278(2) | 0.0078(7) |
| O2 | 16f | 1 | 0.3715(3) | -0.1257(3) | 0.2686(2) | 0.0099(7) |
| O3 | 16f | 1 | 0.2791(3) | -0.2397(3) | 0.1301(2) | 0.0072(7) |
| O4 | 16f | 1 | 0.1086(3) | -0.0701(3) | 0.0766(2) | 0.0096(7) |
| O5 | 16f | 1 | 0.3608(3) | -0.3768(3) | 0.2541(2) | 0.0119(8) |
| O6 | 16f | 1 | -0.0864(3) | 0.0767(3) | 0.0907(2) | 0.0090(7) |
| La1 | 4b | 0.113 | 1/2 | -1/4 | 1/8 | * |
| Sr1 | 4b | 0.887 | 1/2 | -1/4 | 1/8 | * |
| La2 | 16f | 0.713 | 0.26769(3) | 0.04608(3) | -0.11487(2) | * |
| Sr2 | 16f | 0.287 | 0.26769(3) | 0.04608(3) | -0.11487(2) | * |
| La3 | 8e | 0.378 | 0 | 1/4 | -0.13891(3) | * |
| Sr3 | 8e | 0.622 | 0 | 1/4 | -0.13891(3) | * |
| La4 | 16f | 0.764 | 0.20697(3) | -0.22582(3) | -0.03444(2) | * |
| Sr4 | 16f | 0.236 | 0.20697(3) | -0.22582(3) | -0.03444(2) | * |
| Ir1 | 8c | 1 | 0 | 0 | 0 | * |
| Ir2 | 8d | 1 | 1/4 | -1/4 | 1/4 | * |

*Anisotropic Displacement parameter, see table 3

**Table S3 Anisotropic thermal displacement parameters for $Sr_{4.24}La_{6.76}Ir_4O_{24}$.**

| Atom | U11 | U22 | U33 | U12 | U13 | U23 |
|---|---|---|---|---|---|---|
| La1 | 0.0098(4) | 0.0098(4) | 0.0855(12) | 0.00000 | 0.00000 | 0.00000 |
| Sr1 | 0.0098(4) | 0.0098(4) | 0.0855(12) | 0 | 0 | 0 |
| La2 | 0.00592(17) | 0.00492(18) | 0.00657(16) | 0.00018(12) | 0.00034(12) | 0.00004(12) |
| Sr2 | 0.00592(17) | 0.00492(18) | 0.00657(16) | 0.00018(12) | 0.00034(12) | 0.00004(12) |
| La3 | 0.0119(3) | 0.0077(3) | 0.0092(3) | 0.0036(2) | 0 | 0 |
| Sr3 | 0.0119(3) | 0.0077(3) | 0.0092(3) | 0.0036(2) | 0 | 0 |
| La4 | 0.00720(18) | 0.00552(17) | 0.00894(18) | -0.00138(12) | 0.00003(12) | 0.00044(11) |
| Sr4 | 0.00720(18) | 0.00552(17) | 0.00894(18) | -0.00138(12) | 0.00003(12) | 0.00044(11) |
| Ir1 | 0.00397(14) | 0.00443(14) | 0.00483(15) | -0.00010(9) | -0.00048(9) | 0.00041(9) |
| Ir2 | 0.00361(14) | 0.00335(14) | 0.00394(15) | 0.00049(9) | 0.00032(9) | 0.00011(9) |

**Table S4. Site occupancies for La and Sr used in the hypothetical models for $Sr_xLa_{11-x}Ir_4O_{24}$ in the LMTO calculations, in space group $I\,4_1/a$.**

|  | $Sr_1La_{10}Ir_4O_{24}$ | $Sr_2La_9Ir_4O_{24}$ | $Sr_3La_8Ir_4O_{24}$ | $Sr_4La_7Ir_4O_{24}$ | $Sr_5La_6Ir_4O_{24}$ |
|---|---|---|---|---|---|
| 4b | Sr | La | Sr | La | Sr |
| 8e | La | Sr | Sr | La | La |
| 16f-I | La | La | La | Sr | Sr |
| 16f-II | La | La | La | La | La |

**Figure S1. Density of states in LSDA of $SrLa_{10}Ir_4O_{24}$. Compare to Figure 8 to observe that spin-polarization does not split $t_2g$ band**

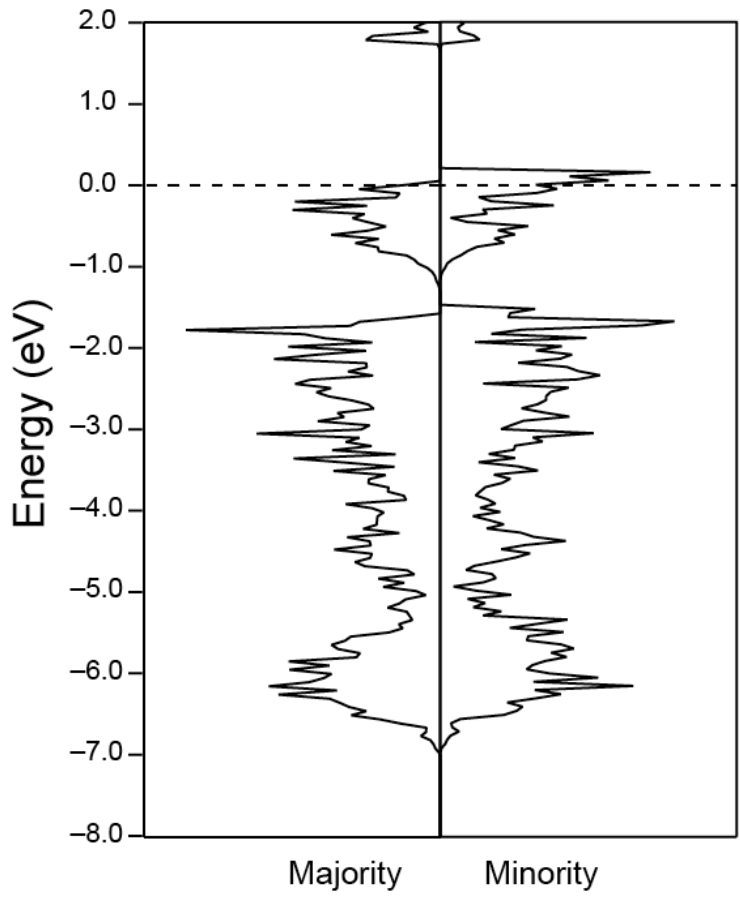